# Impedance-matched High-Overtone Thickness-Shear Bulk Acoustic Resonators with Scalable Mode Volume

Zi-Dong Zhang[1,2] *, Zhen-Hui Qin[1], Yi-Han He[1], Yun-Fei Cheng[1], Hao Yan[1], Si-Yuan Yu[1,3,4], Ming-Hui Lu[1,3,4] †, and Yan-Feng Chen[1,3]

[1] State Key Laboratory of Solid State Microstructures, Department of Materials Science and Engineering, Nanjing University, Nanjing 210093, China

[2] Institute for Frontier Science, Nanjing University of Aeronautics and Astronautics, Nanjing 210016, China

[3] Collaborative Innovation Center of Advanced Microstructures, Nanjing University, Nanjing 210093, China

[4] Jiangsu Key Laboratory of Artificial Functional Materials, Nanjing University, Nanjing 210093, China

Email: zdzhang@nuaa.edu.cn (*Zi-Dong Zhang*); luminghui@nju.edu.cn (*Ming-Hui Lu*)

**Abstract**

High-overtone bulk acoustic resonators are essential components in microwave signal processing and emerging quantum technologies, yet conventional designs face limitations such as limited impedance matching, spurious mode interference, and restricted scalability. Here, we introduce a laterally excited high-overtone thickness-shear bulk acoustic resonator (X-HTBAR), which overcomes these limitations through a planar excitation scheme. The X-HTBAR employs a 3 μm 128° Y-cut LiNbO₃ piezoelectric layer and a 500 μm high-resistivity silicon substrate, enabling efficient excitation of thickness-shear modes via lateral electrodes without requiring bottom electrodes and confining the acoustic field between top electrodes. This configuration eliminates parasitic losses, enhances energy transfer efficiency (>99%), and achieves a stable free spectral range (≈ 5.75 MHz) with minimal fluctuations. Experimental characterization reveals comb-like phonon spectra spanning 0.1–1.8 GHz, high quality factors ($Q > 10^3$–$10^5$), frequency-quality product values ($f \times Q$) exceeding $10^{13}$ at room temperature, and a low temperature **coefficient of frequency (TCF). Moreover, a gridded electrode design and thickness-shear resonators based on 128° Y-cut LiNbO$_3$—insensitive to electrode spacing and featuring a large electromechanical coupling coefficient—effectively suppress spurious modes and enable tunable mode volumes (0.008–0.064 mm³). Together, these features collectively endow X-HTBARs with excellent integration compatibility and immunity from electrode perturbations, positioning them as promising multimode phonon sources for large-scale quantum interconnects and microwave photonic integrated circuits.**

## Introduction

High-overtone bulk acoustic resonators (HBARs), as key components in microwave engineering, have been widely adopted in radio-frequency signal processing[1,2], low phase-noise reference sources[3], and materials characterization systems[4,5], owing to their excellent scalability, high-frequency broadband characteristics, superior quality factors, and versatile control over multiphonon modes[6-13]. In recent years, HBARs have attracted increasing attention for their breakthrough applications in microwave-to-optical conversion[14-16] and quantum information technologies[17-20]. Their unique multiphonon-mode cooperative transduction mechanism offers significant potential for emerging platforms such as frequency-multiplexed qubit interconnects and large-scale microwave photonic integrated circuits, providing innovative solutions for the synergistic advancement of quantum computing and classical communication systems[15,20].

HBARs are a class of acoustic devices based on heterostructures, typically comprising a metal/piezoelectric film/metal sandwich-type transducer integrated with a low-loss substrate, as illustrated in Fig. 1a. The transducer excites longitudinal acoustic waves via the inverse piezoelectric effect (see Figs. 1b and 1c) and couples them into the underlying substrate. Acting as a Fabry–Pérot-type acoustic cavity, the substrate reflects acoustic waves through impedance mismatch at its top and bottom surfaces, thereby forming stable standing wave modes and enabling high-quality-factor resonances along the longitudinal propagation direction. Multiple reflections of acoustic waves within the cavity give rise to high-order overtone modes with narrow linewidths. The piezoelectric layer is commonly made from low-dielectric- and low-acoustic-loss materials such as aluminum nitride (AlN)[12], scandium-doped aluminum nitride (AlScN)[21], Gallium Nitride (GaN)[17], and barium strontium titanate (Ba$_{0.5}$Sr$_{0.5}$TiO$_3$)[22]. To effectively minimize acoustic energy dissipation, substrates are typically selected from low-acoustic-loss materials such as diamond[21], sapphire[12], or silicon carbide (SiC)[6].

However, in conventional HBARs, the presence of a bottom metal electrode between the piezoelectric layer and the acoustic cavity disrupts the continuous acoustic impedance transition, thereby limiting efficient acoustic power transmission. This impedance mismatch induces fluctuations in mode spacing—i.e., instability in the free spectral range (FSR)—which is detrimental to applications such as information storage. Moreover, traditional HBARs are often designed with irregular geometries to suppress spurious vibration modes and parasitic effects such as stray capacitance and inductance[15,17]. As a result, the resonant mode volume is relatively small and difficult to tailor, hindering their scalability and applicability in large-scale integrated quantum and microwave photonic devices[16]. To improve impedance matching, superconducting niobium nitride (NbN) has been explored as a bottom electrode, significantly enhancing the acoustic coupling efficiency between the piezoelectric layer and the substrate[17]. Nevertheless, the superconducting properties of both the top and bottom electrodes rely on cryogenic conditions, limiting their practical deployment. To overcome these limitations, a novel single-top-metal-electrode HBAR structure has been proposed (Fig. 1d), in which a conductive substrate serves dually as the suspended bottom electrode and the acoustic cavity, thereby eliminating the need for a bottom metal layer[23,24]. By judiciously selecting the material pairings—such as AlN or X-cut $LiNbO_3$ thin films on n-type silicon carbide (SiC) substrates—superior acoustic impedance matching can be achieved, significantly reducing FSR fluctuations[24]. The transducer excites shear horizontal (SH) waves[23] or longitudinal acoustic waves[24] [See Supplementary Note 2]. Nonetheless, phonon scattering at the metal-piezoelectric (introduced by the top metal electrode) heterogeneous interface remains inevitable and continues to restrict device performance. In addition, this structure tends to cause spatial dispersion of acoustic energy, as illustrated in Figs. 1e and 1f, where acoustic waves become localized under different electrodes. This leads to a reduction in acoustic energy injection efficiency. Moreover, the single-top-metal-electrode HBAR requires a costly conductive substrate and still fails to enlarge the resonant mode volume, since the excited acoustic waves are confined between the top metal and bottom nonmetal electrodes (similar to those in conventional HBARs), making the mode volume dependent on the top electrode size.

In this study, we demonstrate a laterally excited HBAR (X-HBAR) based on a $LiNbO_3$-on-Si substrate, featuring improved acoustic impedance matching and scalable mode volume. Unlike conventional lateral overmoded bulk acoustic-wave resonators (LOBARs), which rely on localized piezoelectric transducers to excite A0, S0, and S1 modes that propagate across a suspended substrate to form lateral overtones[25-28], the X-HBAR directly excites thickness shear modes via lateral electrodes without requiring bottom electrodes and confining the acoustic field between top electrodes, enabling high-quality-factor resonances along the longitudinal direction. The resonant region of the X-HBAR comprises only the piezoelectric film and a low-loss substrate, offering superior acoustic impedance matching and significantly reduced acoustic energy dissipation. Moreover, the lateral excitation scheme and the 128° Y-cut LiNbO$_3$, with its large piezoelectric coefficient and insensitivity to electrode spacing, enable a scalable resonant mode volume, thereby improving energy storage and coupling efficiency.

## Results

### X-HTBARs Structural Design and Working Mechanism

HTBARs consist of an electrode/piezoelectric transducer formed on a relatively thick, low-loss substrate, as illustrated in Fig. 1g. The piezoelectric layer employs a 3 μm-thick 128° Y-cut LiNbO$_3$ thin film, while the substrate is a 500 μm-thick (100)-oriented high-resistivity silicon (HR-Si) wafer with a resistivity exceeding 10,000 Ω·cm. The 128° Y-cut LiNbO$_3$ features a large piezoelectric coefficient $e_{15}$ (4.47C/m$^2$)[29], enabling a high effective electromechanical coupling coefficient $k^2$, thereby significantly enhancing the excitation efficiency of thickness-shear modes. Importantly, the piezoelectric coefficients $e_{11}$, $e_{12}$, $e_{13}$, and $e_{14}$ are all zero in this crystal orientation, and the transverse shear coefficient $e_{16}$ is relatively small (0.28 C/m²), effectively suppressing lateral spurious modes such as symmetric Lamb waves and their overtones. This results in enhanced spectral purity of the primary mode. The HR-Si substrate offers multiple advantages: its high resistivity minimizes signal loss and parasitic capacitance, improving high-frequency performance, signal isolation, and integrity; its excellent thermal conductivity supports efficient heat dissipation; and it is low-cost and compatible with standard CMOS processes, offering outstanding process scalability. As shown in Figs. 1h and 1i, the working principle of this structure is as follows: the transducer excites thickness-shear modes via the inverse piezoelectric effect and couples the acoustic energy into the substrate. As further shown in Supplementary Fig. S1, the excitation of longitudinal acoustic waves is significantly weaker than that of thickness-shear waves. Acting as a Fabry–Pérot-type acoustic cavity, the HR-Si substrate reflects acoustic waves through impedance mismatch at its top and bottom surfaces, thereby forming stable standing wave patterns and enabling narrow-linewidth, high-overtone resonances.

An idealized un-attached piezoelectric transducer exhibits a spectral response that can be expressed as[29-31]

$$f_s^{mn} = \sqrt{\left(\frac{mv_z}{2t}\right)^2 + \left(\frac{nv_x}{2p}\right)^2} \quad (1)$$

where $f_s^{mn}$ denotes the resonant frequency of the $(m, n)$ mode, $v_z$ and $v_x$ are the acoustic velocities along the shear and longitudinal directions, respectively. For lithium niobate, these values are 3592 m/s and 6541 m/s, respectively[32]. The variables $t$ and $p$ refer to the thickness of the piezoelectric film and the period of the electrode structure, while $m$ and $n$ are the mode orders along the shear and longitudinal directions, respectively, taking integer values of 1, 2, 3, and so on. This study focuses on antisymmetric modes with a longitudinal mode order of $n = 1$, and shear mode orders of $m = 1$ and $m = 3$. Under the condition that $n = 1$ and the electrode pitch is much larger than the film thickness (i.e., $t/p < 0.1$), the resonant frequencies of the thickness-shear modes can be approximated as

$$f_0^{m1} \approx \frac{v_z}{2t} m. \quad (2)$$

As indicated by Equation (2), for a given mode order, thinner piezoelectric films correspond to higher resonant frequencies. Taking the first-order antisymmetric (A1) mode and third-order antisymmetric (A3) mode as examples, their estimated resonant frequencies are approximately 0.598 GHz and 1.796 GHz, respectively. Figure 2a shows the simulated admittance spectrum of the piezoelectric transducer. The insets illustrate the displacement mode profiles at both the resonant and anti-resonant frequencies. The simulation results reveal that only odd-order antisymmetric modes are effectively excited, with distinct resonant responses observed for the A1 and A3 modes at 0.54 GHz and 1.609 GHz, respectively.

When the resonator is attached to a substrate with finite thickness, it exhibits a comb-like phonon spectrum, as shown in Fig. 2b, where a periodic frequency spacing between modes can be observed. Figure 2b presents the simulated admittance curve of a HBAR, revealing two types of high overtones corresponding to the A1 and A3 modes. The simulated displacement mode shapes of the A1 and A3 overtones are shown in Fig. 2c, confirming that the piezoelectric thin film excites the corresponding A1 and A3 modes and injects vibrational energy into the silicon substrate, thereby forming a high-overtone bulk acoustic resonator. The resonance frequency of the $p$th-order mode of the HBAR is given by

$$f_p \sim p \times \left(\frac{v_{si}}{2t_{si}}\right) \quad (3)$$

where $v_{si}$ and $t_{si}$ denote shear acoustic velocity and thickness of the substrate, respectively. The FSR between consecutive overtone modes is inversely proportional to the substrate thickness. For silicon, the acoustic velocities in the shear and longitudinal directions are 5846 m/s and 8442 m/s, respectively[32]. When the silicon substrate thickness is 500 µm, the FSR is approximately 5.846 MHz. As shown in Supplementary Fig. 3 (a magnified view of Fig. 2c), the FSR extracted from finite element simulations is about 5.8 MHz, which is in good agreement with the theoretical value calculated using equation (3).

Acoustic impedance matching is crucial for efficient energy transfer from the piezoelectric layer to the acoustic cavity[7]. To minimize reflection losses and enhance energy transmission efficiency, the acoustic impedance of the piezoelectric layer must be matched to that of the substrate material. When a thickness-shear mode wave propagates across the interface between two materials, the fractional reflected and transmitted power, denoted as $R$ and $T$, respectively, can be calculated using equation (4)[7].

$$R = \left|\frac{Z_1 - Z_2}{Z_1 + Z_2}\right|^2, \quad T = 1 - R \quad (4)$$

Here, $Z$ denotes the acoustic impedance of the material, defined as $Z = \rho \cdot c$, where $\rho$ and $c$ are the density and the shear acoustic velocity, respectively. The acoustic power transmission ratio from the piezoelectric layer (layer 0) to the $i$th layer is calculated by the following expression[7]

$$P_i = P_0 \times \prod_1^i T_{(i-1) \to i}. \quad (5)$$

where $P_0$ denotes the generated acoustic power in the piezoelectric layer. The density of $LiNbO_3$ is 4628 kg/m³, while that of silicon is 2329 kg/m³. Based on these values, the acoustic impedance for the thickness-shear mode is calculated as 16.62 MRayls for lithium niobate and 13.62 MRayls for silicon, yielding an impedance ratio $Z_t/Z_{sub} = 1.22$. The good impedance matching between $LiNbO_3$ and Si substrate enables highly efficient acoustic power transfer between them, exceeding 99%.

**X-HTBARs Performance Characterization**

We fabricated four types of X-HTBARs with different electrode configurations, as illustrated in Supplementary Figs. 4a–d. Among them, the best-performing devices are based on a two-port single-pair electrode structure and a grid-electrode structure (Supplementary Figs. 4e-h), both equipped with rectangular ground–signal–ground (GSG) electrodes on either side. Their optical microscope images are shown in Fig. 3a and Fig. 3b, respectively. A cross-sectional scanning electron microscope (SEM) image of the LN-on-Si substrate is presented in Fig. 3c, indicating a lithium niobate film thickness of 3.25 μm. Insets in Fig. 3c and Supplementary Fig. 5 present magnified SEM images of the $LiNbO_3$–silicon interface, confirming direct bonding of the $LiNbO_3$ thin film to the silicon substrate. On this heterogeneous substrate, excitation electrodes (200 nm Au/10 nm Cr) were fabricated using electron-beam evaporation followed by photolithographic lift-off, enabling the realization of the X-HTBAR device. The frequency response of the X-HTBAR was characterized in air at room temperature using a vector network analyzer. The admittance curve of the single-pair electrode device is shown in Fig. 3d. The measured results reveal a comb-like phonon spectrum, with resonance modes primarily distributed in two frequency bands: 0.1–0.8 GHz and 1.1–1.8 GHz, corresponding to the A1 and A3 overtones, respectively. The A1 overtones include mode number 113, and the $A_3$ overtones is associated with mode number 114. These experimental results are consistent with the numerical simulations presented in Fig. 2, except for a slight deviation in the impedance ratio. The admittance curve was obtained from a series of devices with varying electrode gaps, with the best-performing device selected for presentation. The optimal device has an electrode gap $g$ of 60 μm, corresponding to a mode volume of 540 μm × 60 μm × 503 μm = 0.0163 mm³. This device exhibits high Q and minimal spurious modes, as highlighted in the magnified view in Fig. 3d. Increasing the electrode area in conventional bulk acoustic resonators reduces resistance, improves power handling, and enhances energy confinement. Motivated by this, we widened the electrodes (Supplementary Fig. 6b), but spurious modes were not eliminated.

To mitigate these spurious modes, we adopted a grid electrode design, in which a large electrode is divided into multiple small segments interconnected by thin busbars, as shown in Fig. 3b. The segmentation breaks the acoustic continuity of a large electrode, effectively suppressing unwanted lateral vibrations. A higher number of grid-electrode pairs increases Q and suppresses spurious modes, whereas smaller duty cycles and narrower electrode gaps further enhance resonator performance. A more detailed explanation is provided in Supplementary Note 4. The ground bus lines are configured to ensure a uniform distribution of the excited acoustic waves across the electrodes, as shown in Supplementary Fig. 7. Taking the above factors into account, the electrode parameters were determined to be $W_e$ = 6 μm, $P_e$ = 10 μm, $D_e$ = 37.5%, and $N_e$ = 24. Furthermore, owing to the properties of 128° Y-cut lithium niobate, which features a large piezoelectric stress constant ($e_{15}$) and renders thickness-shear resonators relatively insensitive to electrode spacing, increasing the active region spacing ($g$) has only a minor effect on the effective electromechanical coupling coefficient while helping to suppress parasitic modes, as shown in Supplementary Fig. 8. Consequently, the proposed HBARs achieve tunable mode volumes while maintaining excellent performance even at large volumes. The admittance curve of the multi-pair electrode device is shown in Fig. 3d, exhibiting a comb-like phonon spectrum primarily distributed over two frequency bands: 0.1–0.8 GHz and 1.1–1.8 GHz. Compared to the single-pair electrode device, this configuration shows virtually no spurious modes across the entire operating bandwidth, as highlighted in the magnified view in Fig. 3d. The mode volume of this device is 320 μm × 100 μm × 503 μm = 0.0161 mm³. Moreover, the active region spacing $g$ can be tuned from 50 μm to 400 μm (corresponding to mode volumes ranging from 0.008 to 0.064 mm³), with all resulting resonance states exhibiting high performance and only a few showing minor parasitic mode perturbations, as shown in Fig. 4a-4c. This demonstrates the excellent scalability of the modal volume in this configuration.

Variations in the FSR offer an effective metric for assessing the acoustic impedance matching between different material layers in HBARs. We extracted the FSR of resonant modes from the device using the relation $\Delta f = f_{p+1} - f_p$ (Fig. 3f). In the current structure, both the A1 and A3 mode resonances exhibit FSRs of approximately 5.75 MHz, which agrees well with the theoretical prediction from equation (3). Statistical analysis further reveals that the average FSR ($\mu_{FSR}$) for the A1 mode is 5.75 MHz with a standard deviation ($\sigma_{FSR}$) of 0.03906 MHz, yielding a normalized FSR variation factor $C_{FSR} = \sigma_{FSR}/\mu_{FSR} = 0.00627$, which reflects the uniformity of mode spacing. For the A3 mode, the mean FSR is also 5.75 MHz, with a standard deviation of 0.033166 MHz and a corresponding $C_{FSR}$ of 0.00577. While these $C_{FSR}$ values are slightly higher than those achieved in some state-of-the-art longitudinal-mode HBARs based on substrates grown by epitaxial processes[21,24], they are expected to decrease significantly by replacing the single-side polished silicon substrate with a double-side polished one or through further optimization of the material stack and device design. Longitudinal-mode HBARs have achieved extremely low $C_{FSR}$ after decades of development, and we believe that the proposed X-HTBARs will similarly attain high performance with continued refinement. The slight variation in FSR endows the HBAR device with the potential to realize a mechanical overtone frequency comb[33]. By performing a fast Fourier transform (FFT) on the admittance curve, the time-domain transmission response of the X-HTBAR was obtained, as shown in Fig. 3g. A periodic pulse sequence with a time interval of approximately $\Delta t \approx 137.91$ ns was observed, corresponding to the round-trip travel time of acoustic phonons within the substrate. It is worth noting that the FSR and $\Delta t$ are inversely related, following $\Delta t = (\Delta f)^{-1}$.

**Quality factors, phonon relaxation time and temperature stability.**

The $Q$ extracted from the measured data for electrode gaps of 100 μm are shown in Fig. 5a. All resonant modes exhibit $Q$ values exceeding $10^3$, with some modes reaching values above $10^4$ near 0.38 GHz. Given the frequency-dependent variation of $Q$, the product of frequency and quality factor ($f \times Q$) is widely adopted as a key figure of merit for evaluating resonator performance, as depicted in Fig. 5b. Most resonant modes exhibit $f \times Q$ products exceeding $10^{12}$. The $Q$ factor can be further tuned by adjusting the spacing $g$ of the active region, as illustrated in Fig. 4. Figs. 5d and 5e show the $Q$ and $f \times Q$ values for the device with an electrode gap of 200 μm, which are significantly enhanced compared with those of the 100 μm-gap device. Notably, the A1 and A3 overtone modes achieve peak Q of $7.19 \times 10^4$ and $9.65 \times 10^3$ at 0.432 GHz and 1.375 GHz, respectively. and $f \times Q$ values of $3.35 \times 10^{13}$ and $1.33 \times 10^{13}$ at 0.517 GHz and 1.375 GHz, respectively. Although these values remain below those achieved by some state-of-the-art longitudinal-mode HBARs, significant performance enhancement is expected through further device optimization—for example, by employing low acoustic-loss substrate materials such as sapphire, or silicon carbide. As shown in Supplementary Table 1, the acoustic impedances corresponding to the shear velocities of these two materials are also well matched with that of LN. Moreover, the phonon relaxation time, calculated as $\tau = 2Q/\omega_m$, is shown in Fig. 5c, with values reaching approximately 19.9 μs. The 200 μm-gap device exhibits a maximum phonon relaxation time of 53.04 μs (Fig. 5f).

To assess the temperature stability of the X-HTBARs, we measured their admittance spectra across a wide temperature range (50–295 K). Figures 5g and 5h display the results for the A1-mode and A3-mode overtones, respectively. The temperature evolution of representative phonon modes— mode #54 near 0.46 GHz (Fig. 5g) and mode #172 near 1.455 GHz (Fig. 5h)—shows an increase in resonance frequency with decreasing temperature (T), due to the substrate's rising shear modulus and the consequent increase in shear sound velocity. Extended data on the $Q_s$ and $Q_p$ trends across multiple overtones are presented in Supplementary Fig. 9. The resonance $Q_s$ remains relatively stable with decreasing temperature but shows a decline at lower temperatures, whereas the anti-resonance $Q_p$ continuously decreases with decreasing temperature. This is attributed to the significant mismatch in the coefficient of thermal expansion (CTE) between Si and LiNbO$_3$, which induces substantial thermal stress in the LiNbO$_3$ layer—especially across large temperature variations such as from room temperature to cryogenic conditions or during high-temperature reflow processes. This issue can be mitigated by: (1) introducing CTE-matched or graded buffer layers (e.g., SiO$_2$, amorphous silicon)[34,35]; (2) reducing the LiNbO$_3$ layer thickness (e.g., < 1 μm) to lower thermal stress and enhance flexibility; (3) replacing hydrophilic direct bonding with surface-activated bonding to minimize residual stress from high-temperature processes. As shown in Supplementary Note 8, a SiO$_2$ compensation layer can be introduced between the LiNbO$_3$ layer and the Si substrate. When its thickness is kept below the acoustic wavelength, it does not degrade the resonator performance, slightly enhances the Q-factor, and only shifts the resonance frequency. Therefore, the temperature dependence of the *Q* factor is not discussed in detail here. Figure 5i shows the fitted temperature dependence of the resonance frequencies for mode #54 and mode #172, corresponding to the A1-mode and A3-mode overtones, respectively. The extracted TCF are approximately –10.395 ppm/K for the A1-mode and –12.082 ppm/K for the A3-mode, which are superior to those of other LiNbO$_3$-based HBARs[23] and significantly lower than most previously reported micromechanical resonators.

    The performance of the X-HTBAR proposed in this work was compared with that of previous HBARs, as summarized in Table 1. Due to material availability, the experiments were conducted on single-side polished wafers, whose non-uniform backside induces phonon diffraction and thereby limits device performance. Nevertheless, comparable *Q*, *f*×*Q* products, and *C$_{FSR}$* were achieved even with large-volume modes—tens to hundreds of times greater than those in previous HBARs. HR-Si preserves the mechanical robustness and fabrication compatibility of standard silicon, while offering significantly lower cost compared to other low-loss substrates such as SiC, sapphire, and diamond. Notably, Intel has adopted HR-Si in quantum research to leverage the mature silicon fabrication ecosystem[36]. Therefore, the X-HTBAR developed in this work represents a promising platform for

silicon-based superconducting circuits, serving as a multimode phonon source that enables frequency combs of coherent mechanical modes for strong coupling with superconducting or spin qubits.

**Conclusions**

This study demonstrates a novel X-HTBAR based on a LiNbO$_3$-on-Si platform. By leveraging the superior piezoelectric properties of 128° Y-cut lithium niobate and the excellent acoustic impedance matching with high-resistivity silicon substrates, the X-HTBAR achieves highly efficient excitation and coupling of thickness-shear modes with minimal energy dissipation. The lateral excitation scheme enables scalable resonant mode volumes while maintaining high spectral purity and suppressing spurious modes, addressing key limitations of conventional HBAR designs. Experimental characterization confirms the realization of stable high-quality-factor resonances with frequency-quality factor products exceeding $10^{13}$, consistent free spectral ranges that align well with theoretical predictions, scalable mode volume, and a low TCF. Such a macroscopic resonator exhibits exceptional performance, even surpassing mechanical bound states in the continuum achieved with phononic crystals[37, 38], while also enabling cooperative interactions among multiple phonon modes. Moreover, the X-HTBAR employs a lateral excitation scheme and a 128° Y-cut LiNbO$_3$ layer, which features a large piezoelectric coefficient and provides relative insensitivity to electrode spacing. These properties enable a scalable resonant mode volume, enhance energy storage and coupling efficiency, and confine the excited acoustic field between electrodes. Collectively, they endow X-HTBARs with excellent integration compatibility and robustness against electrode perturbations, positioning them as promising multimode phonon sources for large-scale quantum interconnects[39] and microwave photonic integrated circuits[40,41], fully compatible with current silicon-based semiconductor processes. Further optimization, such as employing ultra-low-loss substrates like sapphire or silicon carbide, may enhance device performance and offers potential for applications in emerging quantum and classical communication systems.

**Methods**

***Simulations of the HBAR with Different Configurations***

Finite element analysis is carried out using the Piezoelectric Devices Module in COMSOL Multiphysics. We model the spectral response of an idealized lithium niobate piezoelectric transducer suspended without attachment to a substrate, as illustrated in Fig. 2a. This essentially represents a film bulk acoustic resonator with free-free mechanical boundary conditions. This is intended to verify the operating mechanism of the X-HTBARs. In the simulations of different types of HBARs, the electrical configurations are set according to Figs. 1a, 1d, and 1g. The top and bottom boundaries are defined as free boundaries, while the left and right boundaries are set as low-reflection boundaries. The material parameters are taken from the COMSOL material library. During meshing, the maximum element size is set to approximately one-tenth of the wavelength.

*Device Fabrication and Characterization*

The X-HTBARs device was manufactured using a 128° Y-cut lithium niobate-on-silicon substrate. The LN-on-Si substrate was custom ordered from Jingzheng Corporation (NanoLN), China. First, the $LiNbO_3$ was directly bonded onto a HR-Si substrate, followed by polishing the $LiNbO_3$ layer down to 3 μm. Due to fabrication variations, the actual $LiNbO_3$ thickness used in this work is 3.25 μm. Note: Due to material availability, the experiments were performed on single-side polished wafers. Next, the metal electrodes were formed through e-beam evaporation metal and then the lift-off process. The electrode structure consisted of a bilayer configuration (10 nm Cr/200 nm Au). Radiofrequency characterization involved collecting two-port scattering parameters under standard ambient conditions using a calibrated vector network analyzer (Agilent M9005A, 10 MHz-26 GHz range). Electrical contact establishment employed TITAN T26A-GSG0200 probes featuring 200 μm spacing between Ground-Signal-Ground contacts, with the device under test positioned on a probe station platform. The input power level from the network analyzer was maintained at -5 dBm during all measurements. The S parameter test under vacuum and cryogenic temperature is realized by a closed-cycle vacuum cryogenic temperature probe station (Inplatinum Scientific Instruments (Shanghai) Co., Ltd.-CPS-50) connected with vector network analyzer (Agilent M9005A, 10 MHz-26 GHz range), where the room temperature vacuum is $10^{-4}$ mbar and the cryogenic temperature vacuum is about $5\times10^{-6}$ mbar.

**Data availability**

The main data supporting the findings of this study are available within this letter and its supplementary information.

**Acknowledgements**

This work is supported by the National Key R&D Program of China (Grants No. 2022YFA1404404), the National Natural Science Foundation of China (Grant No. 52250363, No. 52403377, No. 52322201), the Basic Research Program of Jiangsu Province (BK20244002).

**Author contributions**

Z.-D. Z. conceived the idea. Z.-D. Z. carried out the sample fabrication and room-temperature measurements. Z.-D. Z., Z.-H. Q. and Y.-H. He. performed the low-temperature measurements. All authors contributed to data analysis and manuscript preparation.

**Competing interests**

The authors declare no competing interests.

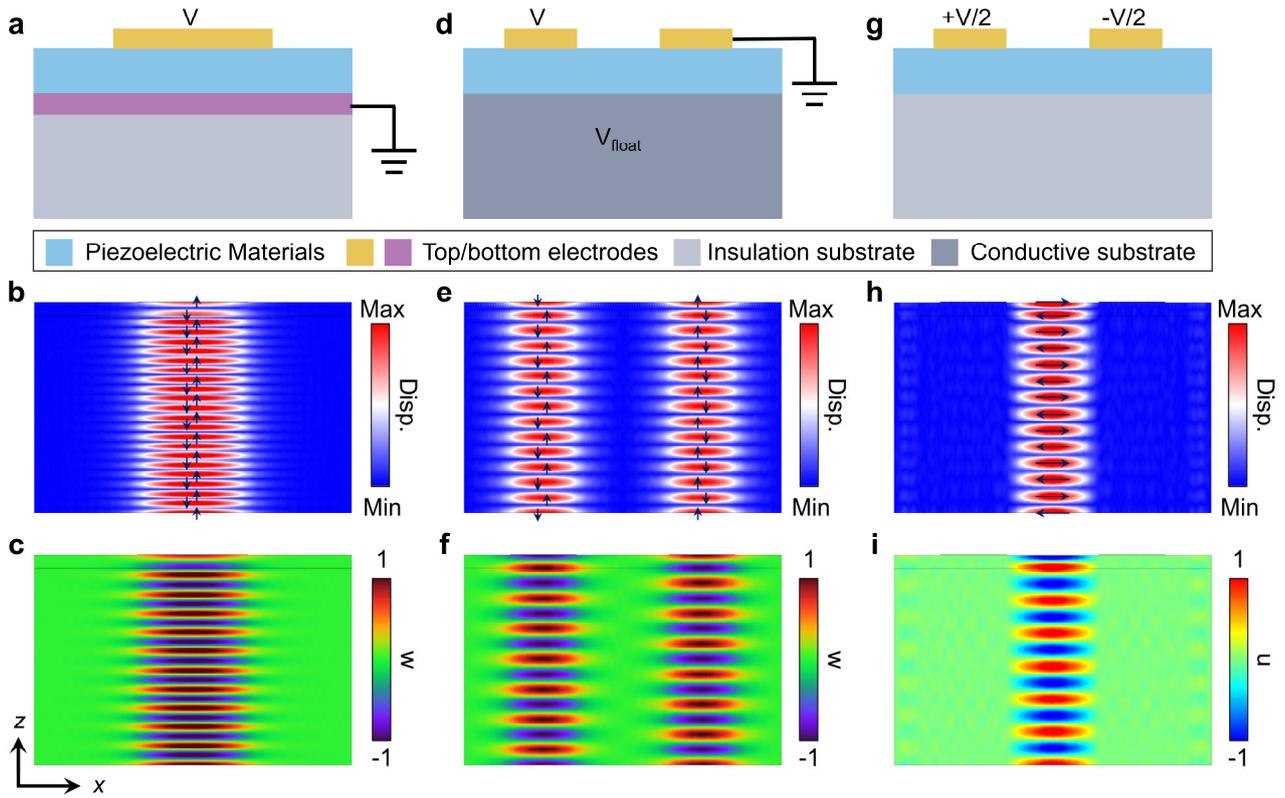

**Fig. 1 Overview of various HBAR configurations. a** Schematic of a conventional HBAR, comprising a metal/piezoelectric/metal transducer stack integrated with a low-acoustic-loss substrate. **b, c** Simulated total displacement field (b) and out-of-plane displacement field (c) at the resonant mode of a conventional HBAR. **d** Schematic of a single-top-electrode HBAR, wherein the conductive substrate simultaneously serves as the suspended bottom electrode and acoustic cavity. **e, f** Simulated total displacement field (e) and out-of-plane displacement field (f) at resonance for the single-top-electrode HBAR. **g** Schematic of a HTBAR consisting of an electrode/piezoelectric transducer on a low-loss substrate. **h, i** Simulated total displacement field (h) and in-plane displacement field (i) at the resonant mode of the HTBAR.

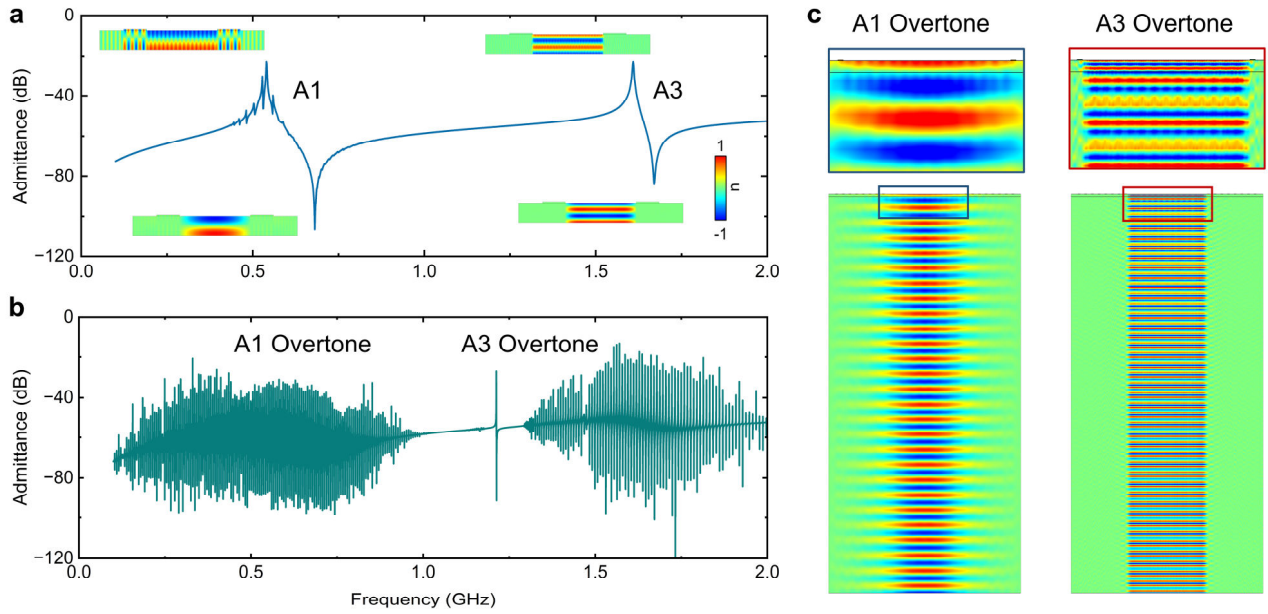

**Fig. 2 Working mechanism of laterally-excited HTBARs. a** Simulated admittance spectrum of an idealized un-attached piezoelectric resonator. Insets illustrate the displacement mode profiles at the resonant and anti-resonant frequencies for the A1 and A3 modes. **b** Simulated admittance spectrum of a laterally-excited HTBAR. **c** Simulated displacement field profiles of the A1 and A3 overtone modes in the laterally-excited HTBAR.

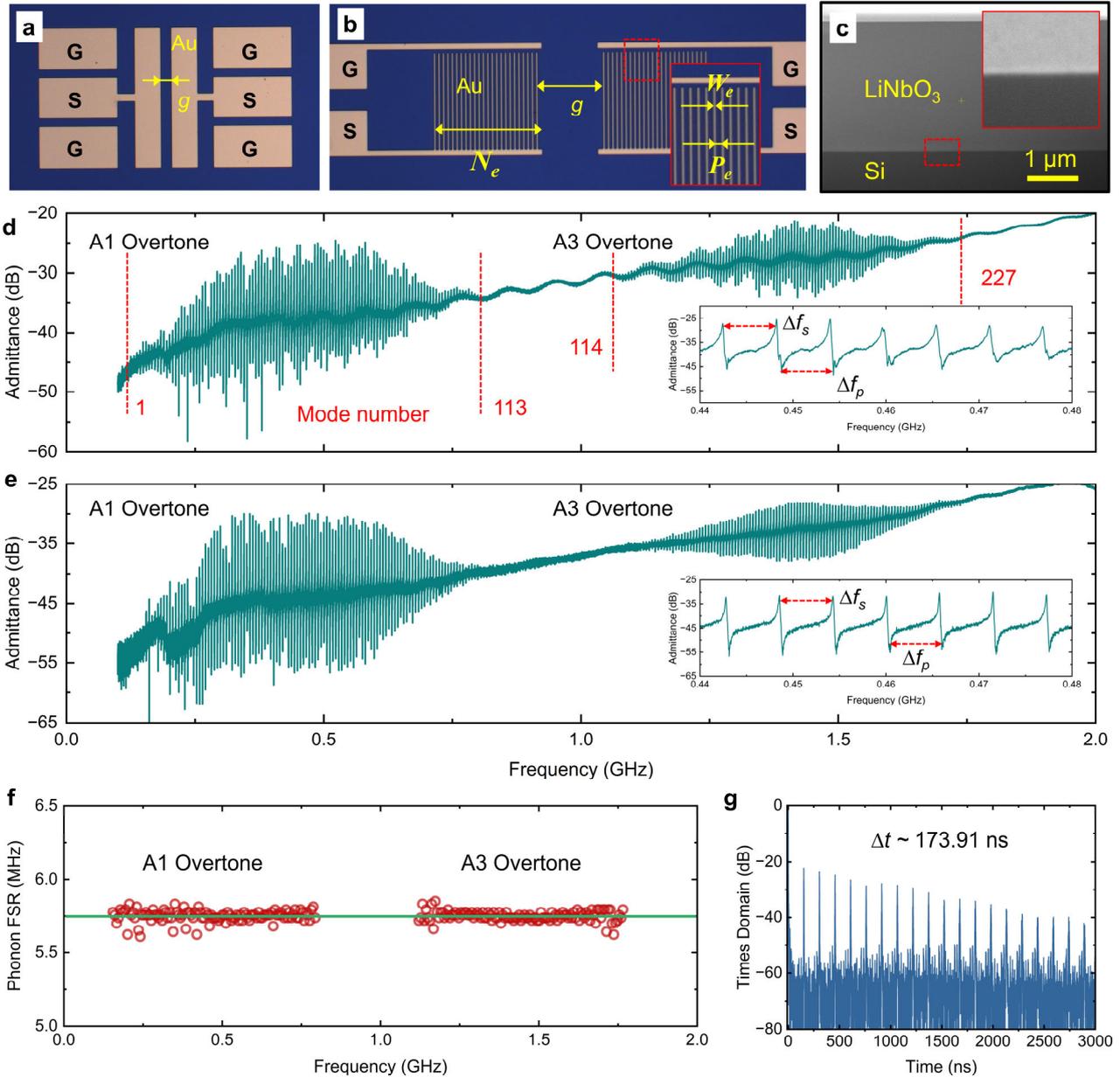

**Fig. 3 Device characterization and microwave measurements of laterally-excited HTBARs at room temperature. a** Optical microscope images of an HTBAR featuring a two-port single-pair electrode configuration. **b** Optical microscope images of HTBARs with a multi-pair electrode design. $W_e$: electrode width, $N_e$: Number of electrode pairs, $P_e$: electrode pitch, $D_e$: electrode duty cycle ($D_e = W_e/(P_e+W_e)$). **c** Cross-sectional SEM image of the LN-on-Si. The inset presents a magnified SEM view of the interface. **d** Measured admittance spectrum of the single-pair electrode device. Inset shows a magnified view highlighting the resonance peaks. **e** Measured admittance spectrum of the multi-pair electrode device. Inset presents an enlarged view of the resonant features. **f** FSR of resonant modes for the laterally-excited HTBARs with a multi-pair electrode. **g** Time-domain response of the laterally-excited HTBARs obtained via fast Fourier transform (FFT) of the admittance spectrum.

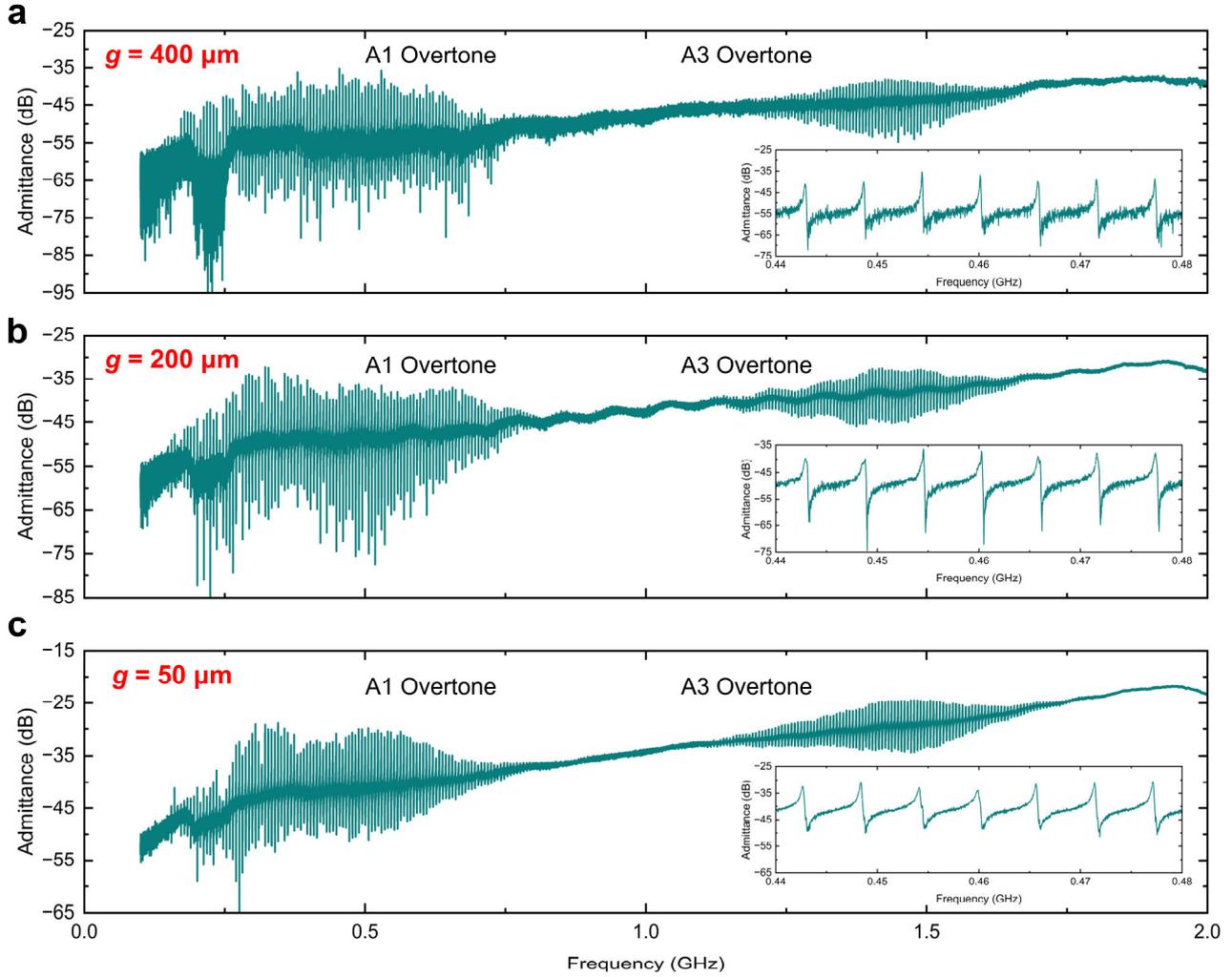

**Fig. 4 Effect of Electrode Gap Spacing on Admittance Spectrum.** Measured admittance spectrum of the multi-pair electrode device with different the active region spacing *g* for 400 μm (a), 200 μm (b), and 50 μm (c), respectively. Inset shows a magnified view highlighting the resonance peaks. As the gap size g varies, it primarily affects the insertion loss, with larger *g* leading to higher insertion loss. Additionally, the Q of the resonator changes with *g*. When g equals 200 μm, the anti-resonance Q reaches its maximum. However, across all values of *g*, no spurious modes appear in the resonance spectra.

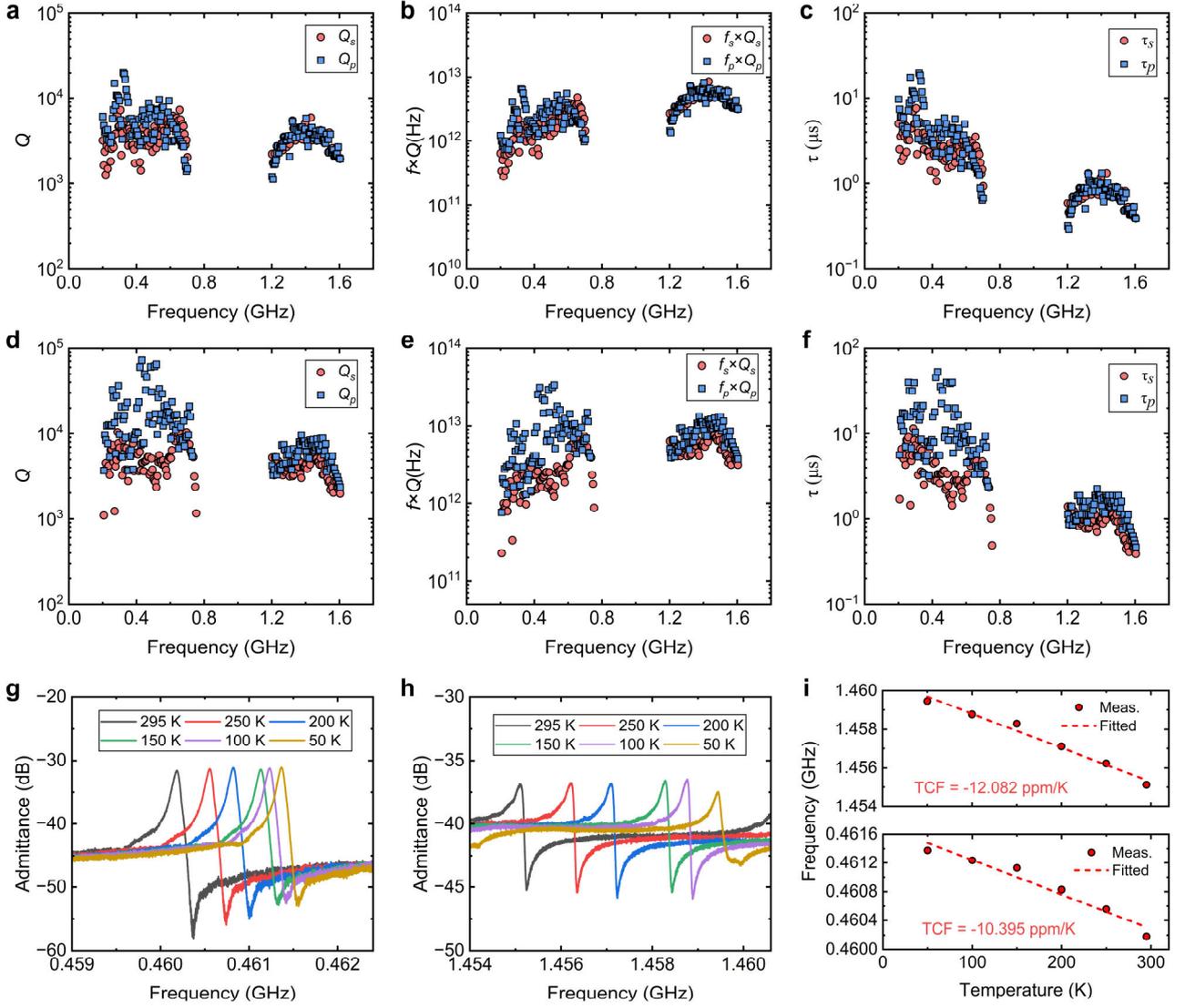

**Fig. 5 Mechanical quality factors, phonon lifetimes and the temperature stability of the X-HTBARs. a, d** Measured mechanical $Q$ for the laterally-excited HTBAR with a multi-pair electrode configuration, for electrode gaps of 100 μm (**a**) and 200 μm (**d**). $Q_s$ is the resonance $Q$, and $Q_p$ is the anti-resonance $Q$. **b, e** $f \times Q$ products for 100 μm (**b**) and 200 μm (**e**) gaps, which extracted from the same device. **c, f** Extracted phonon relaxation times (τ) for 100 μm (**c**) and 200 μm (**f**) gaps of the measured laterally-excited HTBAR. These values were obtained under ambient temperature and pressure conditions. **g**, **h** Admittance spectra of the X-HTBAR measured at various temperatures from 50 K to 295 K, corresponding to the (d) A1-mode overtone and (e) A3-mode overtone, respectively. **i** Temperature-dependent resonance frequencies and the extracted TCF.

Table 1: Comparison of Existing HBARs Configurations

| Material stack | Acoustic mode | Bottom electrode | Excitation region | Cost | $fQ$ ($10^{13}$Hz) | $C_{FSR}$ | Mode V[g] (mm$^3$) | Ref. |
|---|---|---|---|---|---|---|---|---|
| M[a]/GaN/NbN/SiC | L-mode[b] | Yes, M | Top–bot[e] | High | 230 | $8.2\times10^{-4}$ | 0.003 | [17] |
| M/AlN/n-SiC | L-mode | Yes, n-SiC | Top–bot | Medium | 1.3 | $4.8\times10^{-4}$ | 0.00023 | [24] |
| M/X-cut LN/n-SiC | SH-mode[c] | Yes, n-SiC | Top–bot | Medium | 9.6 | | 0.0004 | [23] |
| M/BSTO/M/sapphire | L-mode | Yes, M | Top–bot | Medium | 6 | $3.1\times10^{-3}$ | 0.003 | [5] |
| M/AlScN/M/diamond | L-mode | Yes, M | Top–bot | High | 40 | $2.6\times10^{-3}$ | 0.00023~0.0049 | [21] |
| **M/128°Y-cut LN/HR-Si** | **TS-mode[d] (*A1, A3*)** | **No** | **Top-Top[f]** | **Very low** | **3.35 (A1) 1.33 (A3)** | **$6.27\times10^{-3}$ $5.7\times10^{-3}$** | **0.008~ 0.064** | **Ours Ours** |

[a]M: metal, [b]L-mode: longitudinal-mode, [c]SH-mode: shear-horizontal mode, [d]TS-mode: thickness-shear-mode. [e]Top–bot: Acoustic wave excited between top and bottom electrodes. [f]Top-Top: Acoustic wave excited between top electrodes. [g]V: volume.